\documentclass[twocolumn,nofootinbib,superscriptaddress]{revtex4-1}
\usepackage{latexsym,epsfig,amssymb, amsmath,nicefrac}

\usepackage{color}

\usepackage[makeroom]{cancel}
  \usepackage{latexsym}
  \usepackage{epsf}
  \usepackage{amssymb}
  \usepackage{graphicx}
  \usepackage{verbatim}

\usepackage{graphicx}
\usepackage{dcolumn}
\usepackage{bm}
\usepackage{color}
\usepackage{hyperref}
\usepackage{amsmath}
\usepackage{slashed}

\newcommand{\black}{\color{black}}

\newcommand{\be}{\begin{equation}}
\newcommand{\ee}{\end{equation}}
\newcommand{\citeseq}{\cite{seq1,seq2,seq3,seq4,seq5,seq6,seq7,seq8,seq9,seq10}}
\newcommand{\citeseqshort}{\cite{seq1,seq4,seq6,seq8,seq9, seq10}}

\makeatletter
\renewcommand*\env@matrix[1][\arraystretch]{%
  \edef\arraystretch{#1}%
  \hskip -\arraycolsep
  \let\@ifnextchar\new@ifnextchar
  \array{*\c@MaxMatrixCols c}}
\makeatother

\begin{document}


\title{The Super-St\"uckelberg procedure and dS in Pure Supergravity }

\author{Silvia Nagy} 
\email{ silvia.nagy@nottingham.ac.uk}
\author{Antonio Padilla} 
\email{antonio.padilla@nottingham.ac.uk}
\affiliation{School of Physics and Astronomy, 
University of Nottingham, Nottingham NG7 2RD, UK}
\author{Ivonne Zavala}
\email{ e.i.zavalacarrasco@swansea.ac.uk}
\affiliation{Department of Physics, Swansea University, Singleton Park, Swansea, SA2 8PP, UK}

\date{\today}

\begin{abstract}
Understanding de Sitter space in supergravity - and string theory - has lead to an intense amount  of work for more than two decades, largely motivated by the discovery of the accelerated expansion of the universe in 1998. In this paper, we consider a non-trivial generalisation of  unimodular gravity to minimal ${\cal N}=1$ supergravity, which allows for de Sitter solutions without the need of introducing any matter. We formulate a superspace version of the St\"uckelberg procedure, which restores diffeomorphism and local supersymmetry invariance. This introduces the goldstino associated to spontaneous breaking of supersymmetry in a natural way.
The cosmological constant and gravitino mass are related to the vacuum expectation value of the components of a Lagrange multiplier imposing a super-unimodularity condition.  \\

\textit{Dedicated to Mike Duff on the occasion of his 70th birthday.}

\end{abstract}

\pacs{Valid PACS appear here}

\maketitle


\section{Introduction}

One of the most challenging problems in fundamental physics today is the surprising discovery about twenty years ago \cite{Perlmutter:1997zf,Riess:1998cb}, that the Universe's expansion is currently accelerating. According to the ``standard model" of cosmology, the $\Lambda$CDM ($\Lambda$-Cold-Dark-Matter) model, this acceleration is driven by a tiny constant energy density, $\Lambda$,  whose finely tuned value is the subject of the so-called cosmological constant problem \cite{Zel, Wein, Pol,  Cliff, Martin, me}.    Regardless of how this fine tuning is resolved, be it through anthropic arguments \cite{Wein-anth, Davies-anth, Barrow-anth} or some other mechanism (see e.g \citeseqshort), observational evidence seems to suggest that our universe is asymptotically de Sitter (dS) with a tiny vacuum energy. This has  led to an intense theoretical activity to construct dS vacua in string theory and supergravity (see \cite{Danielsson:2018ztv} for a recent review). 

In the context of supergravity, a long-standing question has been the possibility of finding dS solutions in the pure $\mathcal{N}=1$ model (the simplest supergravity theory). The difficulty is related to the absence of scalar fields (which only appear in matter supermultiplets in $\mathcal{N}=1$, see \cite{Wess:1992cp}). In string theory, the KKLT construction of de Sitter vacua proposed in 2003 \cite{KKLT} led  to the so called landscape (see \cite{Bousso:2012dk} for a review). However, 
the consistency of these solutions has recently been called into question, so much so that stable de Sitter vacua are conjectured to be part of the string theory swampland  (see \cite{Brennan:2017rbf,Danielsson:2018ztv,Palti:2019pca} for recent reviews and  references). Given the observational evidence in favour of a small and positive vacuum energy, the proof or disproof of this conjecture is of crucial importance. Although our immediate focus is on supergravity,  one of the long term motivations for our work is to open up new ideas for seeking de Sitter vacua in string theory.

 Returning to KKLT,  one of  its key ingredients  is the uplift from a supersymmetric AdS vacuum to a dS vacuum via the introduction of an anti-D3-brane.  Recent developments have made good progress in clarifying the role of the anti-D3-brane in providing de Sitter vacua with spontaneously broken supersymmetry in 
 terms of a 4D, ${\cal N}= 1$ supersymmetric action \cite{Ferrara:2014kva,Kallosh:2014wsa,Bergshoeff:2015jxa}.  
It corresponds to a globally supersymmetric Volkov-Akulov (VA) Goldstino theory \cite{Volkov:1972jx,VOLKOV1973109} coupled to a supergravity background. 
The crucial ingredient in this approach is the use of a nilpotent constrained superfield,  which is a regular chiral superfield, $X$,  that  satisfies the constraint, $X^2=0$. This constraint eliminates the scalar component of the chiral multiplet, which  requires that supersymmetry be non-linearly realised\footnote{Constrained superfields  has  proved important for building cosmological and phenomenological models in the framework of supergravity and string theory, see e.g.~\cite{Kallosh:2015nia,Garcia-Etxebarria:2015lif,Dasgupta:2016prs,Kallosh:2016aep,Aalsma:2017ulu,Cribiori:2018dlc,Kallosh:2018psh,Kallosh:2018nrk,Kallosh:2017wnt,GarciadelMoral:2017vnz,Vercnocke:2016fbt,Cribiori:2019hod,Antoniadis:2014oya,Ferrara:2014kva,Kallosh:2014via,DallAgata:2014qsj,Kahn:2015mla,Ferrara:2015tyn,Carrasco:2015iij,DallAgata:2015zxp,Dudas:2016eej,Kallosh:2016ndd,McDonough:2016der,Hasegawa:2017hgd,Farakos:2018sgq,Dalianis:2017okk,Hasegawa:2015era,Linde:2016bcz,Antoniadis:2010hs,Farakos:2012fm,Goodsell:2014dia}. }.  
A complete local supergravity action with  non-linearly  realised  supersymmetry  using constrained superfields was  developed in \cite{Dudas:2015eha,Bergshoeff:2015tra,Hasegawa:2015bza,Ferrara:2015gta,Antoniadis:2015ala,DallAgata:2015pdd,Bandos:2015xnf,Farakos:2016hly,Cribiori:2016qif,Bandos:2016xyu,Lindstrom:1979kq}. These developments allowed for the construction of pure $\mathcal{N}=1$ models which admit dS solutions. In particular, the strategy of \cite{Bergshoeff:2015tra} was  to  introduce a (chiral) Lagrange multiplier, $\Lambda$, in the superconformal theory. In the superconformal action, supersymmetry is linearly realised as long as the Lagrange  multiplier is present. When the equation of motion for $\Lambda$ is solved, it leads to a constraint on the chiral superfield\footnote{See  \cite{Ferrara:2016een} for generalisations.}. For a review of constrained superfields and applications, see \cite{Farakos:2017bxs}.   

In this paper, we take a very different approach, inspired by classical unimodular gravity \cite{ng,Buch1,Buch2,HT, Kuchar, Ellis, Fiol, meUMG}.  Unimodular gravity is a gauge fixed version of General Relativity, in which the determinant of the metric is fixed to a constant so that the resulting field equations correspond to the traceless Einstein equations. The equivalence to General Relativity is easily demonstrated by taking the divergence of these equations and making use of the Bianchi identity. This results in the standard Einstein equations sourced by the energy momentum tensor and a cosmological constant. The only subtlety is that the cosmological constant enters as an integration constant.  Although this doesn't help with the cosmological constant problem \cite{meUMG}, it does suggest a new way of thinking about de Sitter solutions.  In unimodular gravity, it turns out that one can identify the cosmological constant, positive or otherwise, with the vacuum expectation value (vev) of a Lagrange multiplier imposing the constraint on the determinant of the metric \cite{HT, meUMG}. In this paper, we explore a similar idea in the context of supergravity, fixing the superdeterminant with a Lagrange multiplier, then identifying  a component of its vev  with a cosmological constant of arbitrary sign.  As with unimodular gravity, super-unimodular gravity can be written in a way that respects an extended gauge symmetry - diffeomorphisms and local  supersymmetry - with the help of a  St\"uckelberg trick. This invariant theory admits de Sitter vacua with  supersymmetry spontaneously broken by the vev of the Lagrange multiplier.  As expected, the theory of fluctuations about these solutions yields a massive gravitino coupled to a goldstino on a de Sitter background (see also \cite{Bergshoeff:2015tra}). The goldstino is trivially identified with a St\"uckelberg field and, in the appropriate limit, we recover the VA action. 

The rest of this paper is organised as follows: in the next section we review unimodular gravity, the role of the cosmological constant and the St\"uckelberg procedure for restoring the full set of diffeomorphisms. In section \ref{sec:superUMG}, we extend these ideas to supergravity, demonstrating explicitly how one can obtain de Sitter vacua thanks to the vev of the Lagrange multiplier.  We further study the theory of fluctuations about these vacua, showing how we recover the theory of a massive gravitino coupled to a goldstino.  We conclude in section \ref{sec:conc} having transferred some of the calculational details to  the appendix.

\section{Unimodular Gravity}\label{Unimodular Gravity}
Unimodular gravity \cite{ng,Buch1,Buch2,HT, Kuchar, Ellis, Fiol, meUMG} is obtained from a restricted variation of the Einstein-Hilbert action with matter, where the determinant of the metric is fixed to a constant value, traditionally taken to be $-1$.  Because of  this restriction, the theory is only invariant under transverse diffeomorphisms rather than the full diffeomorphism group. The resulting field equations correspond to the traceless part of Einstein equations:
\be \label{UMeqn}
R_{\mu\nu}-\frac14 R g_{\mu\nu}=8 \pi G_N \left( T_{\mu\nu}-\frac14 T g_{\mu\nu}\right) \ ,
\ee
where $G_N$ is Newton's constant. Because the vacuum energy contributes to the energy-momentum tensor as a pure trace, it appears as if it will drop out of the field equations, thereby alleviating the cosmological constant problem.  However,  as explained in \cite{Fiol, meUMG}, this is not the case.  Unimodular gravity is locally equivalent to GR. This follows trivially from the fact that the unimodularity condition, fixing the determinant of the metric, corresponds to a local gauge choice.  One can see this explicitly by taking the divergence of \eqref{UMeqn} and then integrating the resulting equation.  This reveals a hidden equation of the form $R+8\pi G_N T=\Lambda$,  where the cosmological constant, $\Lambda$, enters as an integration constant. Radiative corrections to the vacuum energy are contained in the trace of the energy-momentum tensor. As $\Lambda$  is  fixed by the boundary conditions, the asymptotic curvature is as sensitive to these radiative corrections as it is in GR.  

Our interest in unimodular gravity has nothing to do with its failed assault on the cosmological constant problem. Rather, it arises from the way in which the cosmological constant enters the game - as an integration constant - which is qualitatively different to how it enters in GR .  We can unpack this a little further by imposing  the unimodularity condition, $\sqrt{g}=\epsilon_0$ with a Lagrange multiplier in the Einstein-Hilbert action.  The action for unimodular gravity can be written as 
\be 
\label{classic_unim}
\!\!\!S=\frac{1}{16 \pi G_N} \!\!\int d^4x\left[\sqrt{-g}R-2\Lambda(x)\left(\sqrt{-g}-\epsilon_0 \right) \right] +S_m[g_{\mu\nu}, \Psi]
\ee
with $\epsilon_0$ being a constant (traditionally set to unity). Here $\Lambda(x)$ is a Lagrange multipler {\it field}, imposing a local constraint on the determinant of the metric. This version of the Lagrangian is not invariant under the full diffeomorphism group (Diff) but only under a subgroup of transformations called transverse diffeomorphism (TDiff), whose parameter satisfies $\nabla^\mu\xi_\mu=0$. Since the Lagrange multiplier,  $\Lambda(x)$, transforms like a scalar, it is obvious that Diffs are broken by the last term 
$\int d^4x\Lambda\epsilon_0$.  We shall assume that the action for the matter fields, $S_m[g_{\mu\nu}, \Psi]$ is Diff-invariant. 

Variation with respect to the metric yields an equation of motion of the form,
\be \label{UMeqnLam}
R_{\mu\nu}-\frac12 R g_{\mu\nu}=-\Lambda(x) g_{\mu\nu}+8 \pi G_N T_{\mu\nu} \, .
\ee
If we eliminate $\Lambda(x)$, we arrive at the traceless equations given in \eqref{UMeqn}.  Alternatively we can take the divergence of \eqref{UMeqnLam} and combine it  with the Bianchi identity, forcing the Lagrange multiplier to be a constant. Thus we recover the equations of motion for General Relativity with the constant vev of the Lagrange multiplier taking on the role of the cosmological constant. 

\subsection{The St\"uckelberg Procedure}\label{The Stueckelberg Procedure gravity only}
The unimodularity condition imposed through \eqref{classic_unim} amounts to a local {\em gauge choice}, so we can restore the full general coordinate invariance through the St\"uckelberg procedure \cite{Stueckelberg:1900zz} of introducing an extra field, which transforms appropriately. We can do this by performing a general coordinate transformation (gct) $x^\mu\to \hat  x^\mu(x)$. The original part of the Lagrangian is invariant under it, and on the Diff-breaking term we get
\be
\label{new_lagr_term}
2\int d^4x\Lambda(x)\epsilon_0\quad\to\quad 2\int d^4\hat x\hat \Lambda(\hat x)\epsilon_0=2\int d^4x|J|\Lambda (x)\epsilon_0\,,
\ee
where $|J|=\text{Det}\left(\frac{\partial \hat x^\mu}{\partial x^\alpha} \right)$. To perform the St\"uckelberg trick, we take this transformed action and promote the coordinate transformation $\hat x^\mu(x)$ to four new fields  $s^{\mu}(x)$. The new action
\be 
S_{St}=\tfrac{1}{16\pi G_N}\int d^4x\left[\sqrt{-g}R-2\Lambda\left(\sqrt{-g}-\text{Det}\left(\frac{\partial s}{\partial x} \right)\epsilon_0 \right) \right]\,,
\ee
 is now invariant under the full diffeomorphism group, provided that the St\"uckelberg fields $s^\mu(x)$ transform as scalars:
\be
\hat s^\mu\left(\hat x (x)\right)=s^\mu(x).
\ee 
This particular representation of unimodular gravity can be found in \cite{Kuchar} and is also discussed in \cite{meUMG}.  There exists a closely related version of the theory where the St\"uckelberg fields are repackaged in terms of a four-form field strength \cite{HT}. With Diffs fully restored, the constancy of the Lagrange multiplier arises directly from the equations of motion associated with the St\"uckelberg fields. 

Finally, we note that the  St\"uckelberg trick was performed by acting with a \emph{finite} transformation on the coordinates. However, in what follows,  it will prove instructive to see how to restore  diffeomorphism invariance order by order,  by performing an \emph{infinitessimal} transformation either in the passive form:
\be
x^\mu\ \to \ x^\mu+\xi^\mu+\tfrac{1}{2}\xi^\rho\partial_\rho \xi^\mu+...
\ee
or in the active form, where we vary the scalar field $\Lambda$:
\be
\delta\Lambda=-\xi^\rho\partial_\rho\Lambda+\tfrac{1}{2}\xi^\rho\partial_\rho\left(\xi^\mu\partial_\mu\Lambda\right)+... 
\ee
The perturbative form of the St\"uckelberg trick now involves promoting the transformation parameters to fields: $\xi^\mu\to \phi^\mu$. The Lagrangian can then be constructed order by order in the St\"uckelberg field $\phi^\mu$  
\be
\label{Lag_sec_ord}
\begin{aligned}
\mathcal{L}=&\sqrt{-g}R-2\Lambda\sqrt{-g}\\
& + 2\Lambda\epsilon_0\left[1+\partial_\mu \phi^\mu+\tfrac{1}{2}\phi^\rho\partial_\rho\partial_\mu \phi^\mu+\tfrac{1}{2}\left(\partial_\mu \phi^\mu\right)\left(\partial_\rho \phi^\rho\right)+... \right]\,,
\end{aligned}
\ee
and it will be invariant, up to the relevant order, when $\phi^\mu$ transforms as\footnote{Note that we are using $\xi$ again to denote the transformation parameter, after performing the St\"uckelberg trick.}:
\be
\delta \phi^\mu=-\xi^\mu-\tfrac{1}{2}\xi^\rho\partial_\rho \phi^\mu+\tfrac{1}{2}\phi^\rho\partial_\rho \xi^\mu+...  
\ee

\section{Unimodular Supergravity} \label{sec:superUMG}

We  now show how to extend the procedure we discussed in \autoref{The Stueckelberg Procedure gravity only} to the pure $\mathcal{N}=1$ supergravity model, in the ``old minimal" formulation. We will see that this allows for solutions with a cosmological constant of arbitrary sign.  The goldstino field will be included naturally through the St\"uckelberg procedure, and the cosmological constant will appear as a vev in our description.   

We  work with the conventions of   \cite{Wess:1992cp}. The pure supergravity action in chiral superspace with coordinates $X^M=\left(x^\mu,\Theta^\alpha \right)$, is given by
\be
\label{pure_sugra_superfield_action}
S=\tfrac{6}{8\pi G_N}\int d^4x d^2\Theta\mathcal{E} R+h.c. 
\ee
The components of $R$ are given in appendix \ref{Conventions and useful quantities}  and 
\be
\label{volume_density}
\begin{aligned}
\mathcal{E}&=\mathcal{F}_0\ +\ \sqrt{2}\Theta\mathcal{F}_1\ +\ \Theta\Theta \mathcal{F}_2,\qquad\text{with} \\
\mathcal{F}_0&=\tfrac{1}{2}e \,,\\
\mathcal{F}_1&=\tfrac{i\sqrt{2}}{4}e\sigma^\mu\bar{\psi}_\mu \,,\\
\mathcal{F}_2&=-\tfrac{1}{2}eM^*-\tfrac{1}{8}e\bar{\psi}_\mu\left(\bar{\sigma}^\mu\sigma^\nu-\bar{\sigma}^\nu\sigma^\mu \right)\bar{\psi}_\nu\,,
\end{aligned}
\ee
where  $e=\text{det}\,e^{\bf a}_\mu$, with $e^{\bf a}_\mu$ the vielbein, $\psi_\mu$ the gravitino and $M$ the scalar auxiliary field in the old minimal supergravity model. 
Here $\mathcal{E}$ is a chiral density superfield, characterised by the transformation law
\be 
\delta\mathcal{E}=-\partial_N\left[\left(-1\right)^N\eta^N\mathcal{E}\right]\,,
\ee 
where
\be
\left(-1\right)^N= 
\left\{\begin{matrix}
1, & N=\mu\\ 
-1, &N=\alpha 
\end{matrix}\right.
\ee
and
\be
\label{etas_def}
\begin{aligned}
\eta^\mu(\epsilon)&=\Theta^\beta y^\mu_{1\beta}(\epsilon) + \Theta^2 y^\mu_2(\epsilon) \,,\\
\eta^\alpha(\epsilon)&=\epsilon^\alpha+\Theta^\beta\Gamma^\alpha_{1\beta}(\epsilon)+\Theta^2\Gamma^\alpha_2(\epsilon)\,,
\end{aligned}
\ee
where $\epsilon$ is the parameter of local SUSY transformations. For conciseness, we introduced the following notation
\be
\label{y_gamma_notation}
\begin{aligned}
y^\mu_{1\alpha}(\epsilon)=&2i\left(\sigma^\mu\bar{\epsilon}\right)_\alpha\,,\\ 
y^\mu_2(\epsilon)=&\bar{\psi}_\nu\bar{\sigma}^\mu\sigma^\nu\bar{\epsilon}\,,\\
\Gamma^\alpha_{1\beta}\left(\epsilon\right)=&-i\left(\sigma^\mu\bar{\epsilon}\right)_\beta\psi_\mu^\alpha\,,\\
\Gamma^\alpha_2(\epsilon)=&-i\omega_\mu^{\alpha\beta}\left(\sigma^\mu\bar{\epsilon}\right)_\beta
+\tfrac{1}{3}M^*\epsilon^\alpha \\
&-\tfrac{1}{2}\psi_\nu^\alpha\left(\bar{\psi}_\mu\bar{\sigma}^\nu\sigma^\mu\bar{\epsilon}\right) 
+\tfrac{1}{6}b_\mu\left(\varepsilon\sigma^\mu\bar{\epsilon}\right)^\alpha \,,
\end{aligned} 
\ee
where $\omega_\mu^{\alpha\beta}$ is the spin connection and $b_\mu$ is the vector auxiliary field in the old minimal model. A chiral density superfield can thus be thought of as the supersymmetric analogue of the scalar density, and $\mathcal{E}$ in particular is the supersymmetric version of the measure $\sqrt{-g}$. We define the unimodular supergravity action to be:
\be 
\label{unimodular_sugra_original_action}
S=\tfrac{6}{8\pi G_N}\int d^4x d^2\Theta\left[\mathcal{E} R -2\Lambda\left(\mathcal{E}-\mathcal{E}_0\right)  \right]        +h.c.
\ee
where 
\be
\Lambda=\Lambda_0+\sqrt{2}\black\Theta \Lambda_1+\Lambda_2\Theta^2 
\ee
is now a lagrange multiplier chiral superfield, and we defined\footnote{We take the spinor component of $\mathcal{E}_0$ to vanish for simplicity.}
\be
\mathcal{E}_0=\epsilon_0 -\tfrac{1}{2}  \black m\Theta^2 \,,
\ee
 with $\epsilon_0$ and $m$ real constants\footnote{An alternative formulation of unimodular supergravity, in component form, was given in  \cite{Nishino:2001gd}. The model in   \cite{Nishino:2001gd} can be shown to be equivalent to the $m=0$ limit in our approach.}. Varying over $\Lambda$, we get 
\be
\label{contraint_uni_susy}
\mathcal{E}=\mathcal{E}_0,
\ee 
which is the SUSY analogue of the unimodularity condition. In components this condition reads: 
\be
\label{susy_unimodular_constraints}
\begin{aligned}
\tfrac{1}{2}e=&\epsilon_0 \,, \\
\tfrac{i\sqrt{2}}{4}e\sigma^\mu\bar{\psi}_\mu=&0 \,, \\
-\tfrac{1}{2}eM^*-\tfrac{1}{8}e\bar{\psi}_\mu\left(\bar{\sigma}^\mu\sigma^\nu-\bar{\sigma}^\nu\sigma^\mu \right)\bar{\psi}_\nu=&  -\tfrac{1}{2}m \,.
\end{aligned}
\ee 
The action \eqref{unimodular_sugra_original_action} is invariant under a restricted set of SUSY and diffeo transformations, exactly such that they preserve the conditions in \eqref{susy_unimodular_constraints}:
\be
\label{superfield_constraint_uni}
0=\delta\mathcal{E}=-\partial_M\left[(-1)^M\Xi^M\mathcal{E}\right]\qquad \text{TSdiff}
\ee
where 
\be
\label{diff_and_susy}
\Xi^M=\left(\xi^\mu+\eta^\mu(\epsilon),\eta^\alpha(\epsilon)\right) 
\ee
contains both the diffeo and the SUSY parameter. Note that the constraint imposed through \eqref{contraint_uni_susy} amounts to more than a gauge fixing; in this sense, unimodular supergravity is not a naive supersymmetrisation of the unimodular gravity model in \autoref{Unimodular Gravity}. As a consequence, even though the space of solutions of unimodular gravity matches that of standard Einstein gravity with a cosmological constant, this will not be the case for our model, in the sense that we will not be restricted to AdS and flat space backgrounds.

Finally, we impose the following boundary conditions on our Lagrange multiplier superfield:
\be
\label{boundary_cond_L_mult}
\Lambda_0\Big\vert_{\infty}=K_0\,,\qquad
\Lambda_1\Big\vert_{\infty} =0\,, \qquad
\Lambda_2\Big\vert_{\infty}=K_2\,,
\ee
with $K_0, K_2$ some constants.

\subsection{The Super-St\"uckelberg procedure}
The St\"uckelberg trick performed in \autoref{The Stueckelberg Procedure gravity only} can be extended to any local transformation, and here we  extend it to encompass both diffeomorphisms and local supersymmetry transformations\footnote{See \cite{Volkov:1973jd,Volkov:1974ai} for an alternative use of the St\"uckelberg trick in supergravity; in this formulation the action is constructed from invariant one-forms, reminiscent of the original Volkov-Akulov construction \cite{Volkov:1972jx}}. We now perform a transformation $X^M\to Y^M(X)$ to the superspace coordinates\footnote{We will often omit the spinor indices for simplicity.}  on the non-invariant term:
\be 
\begin{aligned}
\int d^4x d^2\Theta 2\black\Lambda(X)\mathcal{E}_0 \quad &\to\\
\int d^4y d^2\Gamma 2\black\Lambda'(Y)\mathcal{E}_0 
&=\int d^4x d^2\Theta 2\black|sJ|\Lambda(X)\mathcal{E}_0\,,
\end{aligned}
\ee
where $|sJ|=\text{Ber}\left(\frac{\partial Y^M}{\partial X^N} \right)$ is the Berezinian\footnote{The Berezinian, $\text{Ber}$, is the generalisation of the determinant to supermatrices - see \eqref{Berezinian_def} for its definition.}. Crucially, unlike in the gravity case, $Y^M$ will \emph{not} be a general superfunction of $X^M$, but will depend on the diffeomorphism and SUSY parameters in a very particular way. Its construction is given perturbatively below. 

At linear order in the parameters, we have the usual transformation
\be
\delta^{(1)} X^M = \Xi^M \,,
\ee
with  $\Xi^M$ defined in \eqref{diff_and_susy} with the notation we introduced in \eqref{y_gamma_notation}. At second order, we have 
\be
\label{2nd_ord_transform}
\delta^{(2)} X^M= \tfrac{1}{2}\Xi^R\partial_R \Xi^M+\tfrac{1}{2}\delta^{(s)}\left(\Xi^M\right)\,,
\ee
where the first term is the usual one at second order, while the second one takes into account the fact that the objects \eqref{y_gamma_notation} appearing in the SUSY transformations of the coordinates are not arbitrary functions, but depend on the supergravity fields $e_\mu, \psi_\mu, b_\mu, M$. The explicit form of these transformations is given in appendix \ref{app:Supergravity Transformations}. We can thus proceed, order by order in the transformation parameters, and write:
\be 
\label{gen_tranf_sup}
X^M\quad\to\quad 
X^M+\delta^{(1)} X^M+\delta^{(2)} X^M+...\equiv Y^M(X,\Xi),  
\ee
Again, we promote the parameters to fields to get:
\be
\begin{aligned}
\qquad\qquad\qquad\quad\left.\begin{matrix}
\xi^\mu\to\phi^\mu\\ 
\epsilon\to\zeta
\end{matrix}\right\} &\Rightarrow \\
\Xi^M=\left(\xi^\mu+\eta^\mu(\epsilon),\eta^\alpha(\epsilon)\right)\quad&\to\quad
\varphi^M=\left(\phi^\mu+\eta^\mu(\zeta),\eta^\alpha(\zeta)\right)\\
Y^M(X,\Xi) \quad&\to\quad \Phi^M(X,\varphi)
\end{aligned} 
\ee
At this point, we can construct the Lagrangian\footnote{Note that, unlike in the gravity case, the symmetry breaking term $\mathcal{E}_0=\epsilon_0  -\tfrac{1}{2}  \black m\Theta^2 $ depends on the superspace coordinates, so it will be transformed when we perform the St\"uckelberg procedure.}:
\be 
\label{full_action_after_stueck}
\!\!\!S=\tfrac{6}{8\pi G_N}\int d^4x d^2\Theta\left[\mathcal{E} R - 2\black\Lambda\!\left(\!\mathcal{E}-
\text{Ber}\!\left(\frac{\partial \Phi}{\partial X}\right) \mathcal{E}_0\left(\Phi\right)\!\right)  \!\right]        +h.c.
\ee
In analogy with \autoref{The Stueckelberg Procedure gravity only}, we  construct the superjacobian matrix as:
\be 
\label{super_jac_explicit}
sJ=\frac{\partial \Phi^M}{\partial X^N}=
\begin{pmatrix}[1.4]
\tfrac{\partial \Phi^\mu}{\partial X^\nu} & \frac{\partial \Phi^\alpha}{\partial X^\nu}\\
\frac{\partial \Phi^\mu}{\partial X^\beta}  & \frac{\partial \Phi^\alpha}{\partial X^\beta}
\end{pmatrix}
=\bigl(\begin{smallmatrix}
A &B \\ 
C &D 
\end{smallmatrix}\bigr)\,.
\ee
 We now need to compute the superdeterminant, i.e. the Berezinian, defined by
\be 
\label{Berezinian_def}
S\equiv \text{Ber}\left(sJ\right)=\text{Det}\left(A-BD^{-1}C \right)\text{Det}^{-1}(D) \, .
\ee
In components, the Berezinian superfield is given by
\be
S=S_0+\sqrt{2}\black\Theta S_1+S_2\Theta^2  \, ,
\ee
where
\be
\label{stuek_superfield_comps}
\begin{aligned}
\!\!\!\!\!\!S_0=&1+\partial_\mu \phi^\mu-Tr(\Gamma_{1}(\zeta))  
+\tfrac{1}{2}\partial_\mu\left[\phi^\mu\partial_\nu \phi^\nu-\phi^\mu Tr(\Gamma_{1}(\zeta))\right]\\
&-\tfrac{1}{2}Tr(\Gamma_{1}(\zeta))\partial_\mu \phi^\mu +\tfrac{1}{2} \phi^\rho\partial_\rho\Big[ i\psi_\mu\sigma^\mu\Big]\bar{\zeta} 
 +\tfrac{1}{3}M^*\zeta^2\\ 
&+\tfrac{2}{3}M\bar{\zeta}^2\,,\\
\!\!\!\!\!\!\sqrt{2} S_{1}=&\partial_\mu y^\mu_{1}(\zeta)+ 2\Gamma_{2}(\zeta)
+\tfrac{1}{2}\partial_\mu\Big[\phi^\mu\partial_\nu y^\nu_{1}(\zeta)+y^\mu_{1}(\zeta)\partial_\nu \phi^\nu \\  &+2\phi^\mu\Gamma_{2}(\zeta)\Big]
+\Gamma_{2}(\zeta)\partial_\mu \phi^\mu \\
& -\tfrac{1}{2}\partial_\mu \left(\phi^\rho \partial_\rho \Big[ 2i\sigma^\mu\Big]\bar{\zeta}-\partial_\rho \phi^\mu y^{\rho}_{1}(\zeta) \right)\\
&  - \phi^\rho\partial_\rho \Big[ -i \varepsilon\omega_\mu\sigma^\mu  \Big]\bar{\zeta}  -\phi^\rho\partial_\rho M^* \tfrac{1}{3}\zeta 
 -\tfrac{1}{6}\phi^\rho\partial_\rho\Big[ b_\mu\varepsilon\sigma^\mu\Big]\bar{\zeta}\,,\\
\!\!\!\!\!\!S_2=&\partial_\mu y^\mu_2(\zeta)
+\tfrac{1}{2}\partial_\mu\Big[\phi^\mu\partial_\nu y^\nu_2(\zeta)+y^\mu_2(\zeta)\partial_\nu \phi^\nu \\
&-\tfrac{1}{2}y^{\mu}_1(\zeta)\partial_\nu y^\nu_{1}(\zeta) 
 -\tfrac{1}{2}y^{\nu}_1(\zeta) D_\nu y^\mu_{1}(\zeta)\\
&-\phi^\rho\partial_\rho \Big[ \bar{\psi}_\nu\bar{\sigma}^\mu\sigma^\nu\Big]\bar{\zeta} +\partial_\rho \phi^\mu y^\rho_2(\zeta) 
  -\tfrac{4i}{3}M^*\zeta\sigma^\mu\bar{\zeta}\\
& -ib_\nu\bar{\zeta}\bar{\sigma}^\mu\sigma^\nu\bar{\zeta}+i b_\nu\bar{\zeta}\bar{\sigma}^\nu\sigma^\mu\bar{\zeta} \black \Big]\,.
\end{aligned} 
\ee
The crucial point here is that $S_0$, $S_1$ and $S_2$ transform like the components of a standard  chiral density superfield (as given in \eqref{density_comps_tranform}), but the St\"uckelberg fields $\zeta$ and $\phi^\mu$ transform non-linearly. The power of the St\"uckelberg procedure is that it automatically constructs $S=\text{Ber}\left(\tfrac{\partial \Phi^M}{\partial X^N}\right) $ in such a way that \emph{fields which transform non-linearly are embedded in the components of a standard chiral superfield}.

Finally, the SUSY transformations of the St\"uckelberg fields $\phi^\mu$ and $\zeta$ are:
\be
\label{stueck_field_transforms}
\begin{aligned}
\delta \phi^\mu&=\tfrac{1}{2}\left[\zeta y^\mu_{1}(\epsilon)-\epsilon y^\mu_{1}(\zeta) \right]\,,\\
\delta\zeta&=-\epsilon+\tfrac{1}{2}\phi^\rho\partial_\rho\epsilon\,.
\end{aligned} 
\ee

Above we performed the St\"uckelberg trick in the passive form. Just as in the gravity case, the invariant action can alternatively be obtained through an active transformation on the components of the Lagrange multiplier superfield, $\Lambda$. In components, the symmetry breaking terms following from \eqref{unimodular_sugra_original_action} are:
\be
2\Lambda_2 \epsilon_0  -m\Lambda_0\black  +h.c.  
\ee
We want to perform the Super-St\"uckelberg trick up to 2nd order, so we need to know the 2nd order transformation of the component of a chiral superfield in curved space. We can work this out from the general rule \cite{Buchbinder:1998qv}:
\be
\Lambda'(X')=\Lambda(X) \,,
\ee
and making use of \eqref{2nd_ord_transform}  we get (working up to second order in the fermions):
\be
\label{active_transf_lambda}
\begin{aligned}
\delta \Lambda_0=& -\xi^\mu\partial_\mu\Lambda_0-\sqrt{2}\epsilon\Lambda_{1} \\
&+\tfrac{1}{2}\xi^\rho\partial_\rho\left(\xi^\mu\partial_\mu\Lambda_0\right)
+\tfrac{1}{2}\epsilon y^\mu_1(\epsilon)\partial_\mu\Lambda_0 \\
&+\tfrac{\sqrt{2}}{2}\xi^\mu\partial_\mu\left(\epsilon\Lambda_1 \right)
+\tfrac{\sqrt{2}}{2}\xi^\mu\epsilon\partial_\mu\Lambda_1
+\Lambda_2\epsilon^2  \,,\\
\delta \Lambda_1=&-\xi^\mu\partial_\mu\Lambda_1-\tfrac{\sqrt{2}}{2}y_1^\mu(\epsilon)\partial_\mu\Lambda_0
-\sqrt{2}\Lambda_2\epsilon \\
&+\tfrac{\sqrt{2}}{4}\xi^\rho\partial_\rho\left(y_1^\mu(\epsilon)\partial_\mu\Lambda_0\right)+\tfrac{\sqrt{2}}{4}y_1^\rho(\epsilon)\partial_\rho\left( \xi^\mu\partial_\mu\Lambda_0\right)\\
&+\tfrac{1}{2}\xi^\rho\partial_\rho\left(\xi^\mu\partial_\mu \Lambda_1\right)+
\tfrac{\sqrt{2}}{2}\xi^\rho\partial_\rho\left(\epsilon\Lambda_2\right)
+\tfrac{\sqrt{2}}{2}\epsilon\xi^\mu\partial_\mu\Lambda_2\\
&+\left[\tfrac{i\sqrt{2}}{2}\xi^\rho\partial_\rho[\sigma^\mu]\bar{\epsilon}-\tfrac{\sqrt{2}}{4}\partial_\rho\xi^\mu y^\rho_1(\epsilon)\right]\partial_\mu\Lambda_0 \,,\\
\delta \Lambda_2=&-\xi^\mu\partial_\mu\Lambda_2+\tfrac{\sqrt{2}}{2}y^{\mu}_1(\epsilon)\partial_\mu\Lambda_{1}-y^\mu_2(\epsilon)\partial_\mu\Lambda_0-Tr(\Gamma_{1}(\epsilon))\Lambda_2\\
&-\sqrt{2}\Gamma_2(\epsilon)  \Lambda_{1}\\
&+\tfrac{1}{2}\xi^\rho\partial_\rho \Big[\xi^\mu\partial_\mu\Lambda_2-\tfrac{\sqrt{2}}{2}y^{\mu}_1(\epsilon)\partial_\mu\Lambda_{1}+y^\mu_2(\epsilon)\partial_\mu\Lambda_0\\
&+Tr(\Gamma_{1}(\epsilon))\Lambda_2+\sqrt{2}\Gamma_2(\epsilon) \Lambda_{1}\Big]
-\tfrac{1}{4}y^{\rho}_1(\epsilon)\partial_\rho\Big[\sqrt{2}\xi^\mu\partial_\mu\Lambda_{1}\\
&+y^\mu_{1}(\epsilon)\partial_\mu\Lambda_0+2\epsilon\Lambda_2\Big]
+\tfrac{1}{2}y_2^\rho(\epsilon)\partial_\rho\left[\phi^\mu\partial_\mu\Lambda_0\right]\\
&+\tfrac{1}{2}Tr(\Gamma_{1}(\epsilon))\left[\xi^\mu\partial_\mu\Lambda_2\right]
+\tfrac{ \sqrt{2}}{2}\Gamma_2(\epsilon)\xi^\mu\partial_\mu\Lambda_{1}\\
&+\tfrac{\sqrt{2}}{4}\Big[  -  \xi^\rho \partial_\rho \Big[ 2i\sigma^\mu\Big]\bar{\epsilon} + \partial_\rho \xi^\mu y^{\rho}_{1}(\epsilon)\Big]\black
\partial_\mu\Lambda_{1a}\\
&-\tfrac{1}{2}\Big[ -\tfrac{1}{2}y^{\nu}_1(\epsilon) D_\nu y^\mu_{1}(\epsilon)\\
& -\xi^\rho\partial_\rho \Big[ \bar{\psi}_\nu\bar{\sigma}^\mu\sigma^\nu\Big]\bar{\epsilon}           
+\partial_\rho \xi^\mu y^\rho_2(\epsilon) - \tfrac{4i}{3}M^*\epsilon\sigma^\mu\bar{\epsilon}\\
& -ib_\nu\bar{\epsilon}\bar{\sigma}^\mu\sigma^\nu\bar{\epsilon}+i b_\nu\bar{\epsilon}\bar{\sigma}^\nu\sigma^\mu\bar{\epsilon}
\Big]\black\partial_\mu\Lambda_0\\
&+\tfrac{1}{2} \Big[\partial_\mu\epsilon y_{1}^\mu(\epsilon)+
    \xi^\rho\partial_\rho\Big[ i\psi_\mu\sigma^\mu\Big]\bar{\epsilon}\\
&+\tfrac{2}{3}M^*\epsilon^2 +\tfrac{4}{3}M\bar{\epsilon}^2 
\Big] \black\Lambda_2 \\
&-\tfrac{\sqrt{2}}{2}\Big[  \xi^\rho\partial_\rho \Big[ i\omega_\mu\sigma^\mu  \Big]\bar{\epsilon}  - \tfrac{1}{3}\xi^\rho\partial_\rho M^* \epsilon \\
& -\tfrac{1}{6}\xi^\rho\partial_\rho\Big[ b_\mu\varepsilon\sigma^\mu\Big]\bar{\epsilon}\Big]\black\Lambda_{1} \,,
\end{aligned} 
\ee
and then send $\xi^\mu\ \to\ \phi^\mu$ and $\epsilon\ \to\ \zeta$ as before. The action can then be constructed pertubatively as:
\be
\label{active_superstueck_action}
\begin{aligned}
\mathcal{L}=&\sqrt{-g}\Big[R-\tfrac{2}{3}M^*M+\tfrac{2}{3}b^\mu b_\mu+\varepsilon^{\mu\nu\rho\sigma}\Big(\bar{\psi}_\mu\bar{\sigma}_\nu\tilde{\mathcal{D}}_\rho\psi_\sigma\\
&-\psi_\mu\sigma_\nu\tilde{\mathcal{D}}_\rho\bar{\psi}_\sigma \Big) \Big]
+\tfrac{1}{2}\sqrt{-g}\Big[-2\Lambda_2+\sqrt{2}i\Lambda_1\sigma^\mu\bar{\psi}_\mu\\
& +2\Lambda_0\left(\bar{\psi}_\mu\bar{\sigma}^{\mu\nu}\bar{\psi}_\nu+M^* \right)\ +h.c.\Big]\\
&+2\left[\Lambda_2+\delta^{(\epsilon\to\zeta,\xi^\mu\to\phi^\mu)}\Lambda_2 \right]\epsilon_0\\
& -m\left[\Lambda_0+\delta^{(\epsilon\to\zeta,\xi^\mu\to\phi^\mu)}\Lambda_0\right]\black  +h.c.  
\end{aligned} 
\ee

 Finally, the full action, up to second order in the St\"uckelberg fields, obtained either through a passive \eqref{full_action_after_stueck} or an active \eqref{active_superstueck_action} transformation is  (setting $\epsilon_0=\tfrac{1}{2}$) is:

\black
\be\label{action}
\begin{aligned}
\!\!\!S=\tfrac{1}{16\pi G_N}\int\Bigg[&\sqrt{-g}\Big[R-\tfrac{2}{3}M^*M+\tfrac{2}{3}b^\mu b_\mu\\
&+\varepsilon^{\mu\nu\rho\sigma}\Big(\bar{\psi}_\mu\bar{\sigma}_\nu\tilde{\mathcal{D}}_\rho\psi_\sigma
-\psi_\mu\sigma_\nu\tilde{\mathcal{D}}_\rho\bar{\psi}_\sigma \Big) \Big]\\
&+\tfrac{1}{2}\sqrt{-g}\Big[-2\Lambda_2+\sqrt{2}i\Lambda_1\sigma^\mu\bar{\psi}_\mu\\
& +2\Lambda_0\left(\bar{\psi}_\mu\bar{\sigma}^{\mu\nu}\bar{\psi}_\nu+M^* \right)\ +h.c.\Big]\\
&+\Big\{\Big[\Lambda_0 \left(\partial_\mu y^\mu_2(\zeta) -m -m\partial_\mu \phi^\mu\right)\\
&-\Lambda_1\Big(\tfrac{\sqrt{2}}{2}\partial_\mu y^\mu_{1}(\zeta)
+\sqrt{2}\Gamma_{2}(\zeta)-\sqrt{2}m\zeta\Big)\\
&+\Lambda_2\left[1+\partial_\mu \phi^\mu -Tr(\Gamma_{1}(\zeta))\right]  \Big]\\
&+\Lambda_2\Big[\tfrac{1}{2}\partial_\mu\left(\phi^\mu\partial_\nu \phi^\nu - \phi^\mu Tr(\Gamma_{1}(\zeta))\right)\\
&-\tfrac{1}{2}\partial_\mu \phi^\mu  Tr(\Gamma_{1}(\zeta))
+\tfrac{1}{2}\partial_\mu\left(\zeta y^\mu_{1}(\zeta)\right)\\
&+\tfrac{1}{2}\phi^\rho\partial_\rho\Big[ i\psi_\mu\sigma^\mu\Big]\bar{\zeta}
+\tfrac{1}{3}M^*\zeta^2+\tfrac{2}{3}M\bar{\zeta}^2\\
&  -m\zeta^2\black\Big]\\
&\  -\tfrac{\sqrt{2}}{2}\Lambda_1\Big[\tfrac{1}{2}\partial_\mu\Big(\phi^\mu\partial_\nu y^\nu_{1}(\zeta)+y^\mu_{1}(\zeta)\partial_\nu \phi^\nu\\
&+2\phi^\mu\Gamma_{2}(\zeta) \Big)
+\Gamma_{2}(\zeta)\partial_\mu \phi^\mu\\
&-\tfrac{1}{2}\partial_\mu \Big(\phi^\rho \partial_\rho \Big[ 2i\sigma^\mu\Big]\bar{\zeta}
-\partial_\rho \phi^\mu y^{\rho}_{1}(\zeta)  \Big)\\
& - \phi^\rho\partial_\rho \Big[ -i\varepsilon\omega_\mu\left(\sigma^\mu  \Big]\bar{\zeta}\right) \\
&-\tfrac{1}{3}\phi^\rho\partial_\rho M^*\zeta -\tfrac{1}{6}\phi^\rho\partial_\rho\Big[ b_\mu\varepsilon\sigma^\mu\Big]\bar{\zeta}\\
& -m\zeta\partial_\mu \phi^\mu -m\partial_\mu\left(\zeta \phi^\mu\right)         \black\Big]\\
&+\tfrac{1}{2}\Lambda_0\partial_\mu\Big[\phi^\mu\partial_\nu y^\nu_2(\zeta)+y^\mu_2(\zeta)\partial_\nu \phi^\nu\\
&-\tfrac{1}{2}y^{\mu}_1(\zeta)\partial_\nu y^\nu_{1}(\zeta)
 -\tfrac{1}{2}y^{\nu}_1(\zeta)D_\nu y^\mu_{1}(\zeta)\\
&-\phi^\rho\partial_\rho \Big[ \bar{\psi}_\nu\bar{\sigma}^\mu\sigma^\nu\Big]\bar{\zeta} 
+\partial_\rho \phi^\mu y^\rho_2(\zeta)\\
& -\tfrac{4i}{3}M^*\zeta\sigma^\mu\bar{\zeta}
 -ib_\nu\bar{\zeta}\bar{\sigma}^\mu\sigma^\nu\bar{\zeta}+i b_\nu\bar{\zeta}\bar{\sigma}^\nu\sigma^\mu\bar{\zeta}\\
& -m\phi^\mu\partial_\rho \phi^\rho +m\zeta y^\mu_{1}(\zeta)\black \Big] 
 +h.c. \Big\} \,.
\end{aligned}
\ee
Note that, upon application of the St\"uckelberg procedure, the boundary conditions \eqref{boundary_cond_L_mult} are modified (to linear order in the St\"uckelberg field) as:
\be 
\begin{aligned}
\Lambda_0-\phi^\mu\partial_\mu\Lambda_0-\sqrt{2}\zeta\Lambda_1\Big\vert_{\infty}=&K_0\,,\\
\Lambda_1-\phi^\mu\partial_\mu\Lambda_1-\sqrt{2}\zeta\Lambda_2+\tfrac{\sqrt{2}}{2}y_1^\mu(\zeta)\partial_\mu\Lambda_0\Big\vert_{\infty}=&0\,,\\
\Big[\Lambda_2-\phi^\mu\partial_\mu\Lambda_2-\tfrac{\sqrt{2}}{2}y^\mu_1(\zeta)\partial_\mu \Lambda_1-\sqrt{2}\Gamma_2(\zeta)\Lambda_1\qquad\ &\\
-Tr(\Gamma_1(\zeta))\Lambda_2
-y_2^\mu(\zeta)\partial_\mu\Lambda_0\Big]\Big\vert_{\infty}=&K_2\,.
\end{aligned}
\ee
One can recover the boundary conditions up to second order in the St\"uckelberg field by making use of \eqref{active_transf_lambda}.

\subsection{dS solutions}
Let us now look at the possible solutions to the equations of motion derived from \eqref{action} for the bosonic fields,   $g^{\mu\nu}$, $M^*$, $b_\mu$, $\phi^\mu$, $\Lambda_0$ and $\Lambda_2$, which are\footnote{We omit the fermion terms and equations of motion because we are seeking a background solution with vanishing fermions.}:
\be
\label{eom_nonpert}
\begin{aligned}
G_{\mu\nu}+g_{\mu\nu}\left[\tfrac{1}{3}MM^*+\tfrac{2}{3}b^\rho b_\rho+{\rm Re}(\Lambda_2-\Lambda_0M^*)\right]&\\
+b_\mu b_\nu&=0\,,\\
-\tfrac{2}{3}M+\Lambda_0&=0\,,\\
b_\mu&=0\,,\\
-\partial_\mu\Lambda_2+m\partial_\mu\Lambda_0-\tfrac{1}{2}\partial_\nu\phi^\nu\partial_\mu\Lambda_2
+\tfrac{1}{2}\partial_\mu\left(\phi^\nu\partial_\nu\Lambda_2\right)&\\
+\tfrac{m}{2}\partial_\nu\phi^\nu\partial_\mu\Lambda_0-\tfrac{m}{2}\partial_\mu\left(\phi^\nu\partial_\nu\Lambda_0\right)+h.c.&=0\,,\\
\sqrt{-g}M^* -m - m\partial_\mu\phi^\mu- \tfrac{m}{2}\partial_\mu\left[\phi^\mu\partial_\nu\phi^\nu\right]&=0\,,\\
-\sqrt{-g}+1+\partial_\mu\phi^\mu+\tfrac{1}{2}\partial_\mu\left(\phi^\mu\partial_\nu\phi^\nu\right)&=0\,.
\end{aligned}
\ee
These admit the solution
\be
\label{background_sol}
\begin{aligned}
g_{\mu\nu}&=\bar{g}_{\mu\nu},\quad\text{with}\quad \sqrt{-\bar{g}}=1\,,\\
M&=m\,,\\
\Lambda_0&=\tfrac{2}{3}m\,,\\
\Lambda_2&=\mathbf{\Lambda}_2,\quad\text{with}\quad \text{Im}(\Lambda_2)=0\,,
\end{aligned} 
\ee
with all other fields vanishing on the background. The cosmological constant is
\be \label{cc}
c.c.= \mathbf{\Lambda}_2-\tfrac{1}{3}m^2\,.
\ee
Thus, our model allows for a cosmological constant of either sign, similar to the results in the constrained superfields literature \cite{Dudas:2015eha,Bergshoeff:2015tra,Hasegawa:2015bza,Ferrara:2015gta,Antoniadis:2015ala,DallAgata:2015pdd,Bandos:2015xnf,Farakos:2016hly,Cribiori:2016qif,Bandos:2016xyu}. However, we stress that in our approach, the cosmological constant appears as the combination of the vev's of the Lagrange multiplier superfield components $\Lambda_0$ and $\Lambda_2$.

\subsection{Perturbative Treatment and Gravitino Mass}
We now perform a pertubative expansion around the background solution \eqref{background_sol}, in order to study the mass of the gravitino:
\be 
\begin{aligned}
g_{\mu\nu}&=\overline{g}_{\mu\nu}+h_{\mu\nu}\,,\\
\sqrt{-g}&=1+\tfrac{1}{2}h+\tfrac{1}{4}\left(\tfrac{1}{2}h^2-h_{\mu\nu}h^{\mu\nu}\right)\,,\\
\psi_\mu^m&=0+\psi_\mu^m\,,\\
b_\mu&=0+b_\mu\,,\\
M&= m+M\,,\\
\Lambda_0&=\tfrac{2}{3}m+\lambda_0\,,\\
\Lambda_1&=0+\lambda_1\,,\\
\Lambda_2&=\mathbf{\Lambda}_2+\lambda_2, \qquad \text{Im}(\mathbf{\Lambda}_2)=0\,,\\
\phi^\mu&=0+t^\mu\,,\\
\zeta&=0+\zeta\,.
\end{aligned}
\ee
Working to second order in the small perturbations, the action becomes:
\be 
\label{pert_action_dS_sugra}
\begin{aligned}
S=\tfrac{1}{16\pi G_N}\int d^4x \Big[&  2\mathbf{\Lambda_2}  -\tfrac{2}{3}m^2 +\tfrac{1}{4}\bar{\nabla}_\mu h_{\rho\lambda}\bar{\nabla}^\mu h^{\rho\lambda}\\
&-\tfrac{1}{2}\bar{\nabla}_\mu h_{\rho\lambda}\bar{\nabla}^\rho h^{\mu\lambda}
+\tfrac{1}{2}\bar{\nabla}_\mu h^{\mu\nu}\bar{\nabla}_\nu h\\
& -\tfrac{1}{4}\bar{\nabla}_\mu h\bar{\nabla}^\mu h -\tfrac{2}{3}M^*M+\tfrac{2}{3}b^\mu b_\mu\\
& +\tfrac{1}{2}\left(\mathbf{\Lambda_2} -\tfrac{1}{3}m^2\right)\left(h^{\mu\nu}h_{\mu\nu}-\tfrac{1}{2}h^2 \right) 
\\
&+\varepsilon^{\mu\nu\rho\sigma}\left(\bar{\psi}_\mu\bar{\sigma}_\nu\mathbf{D}_\rho\psi_\sigma-\psi_\mu\sigma_\nu\mathbf{D}_\rho\bar{\psi}_\sigma \right)\\
& -\tfrac{2}{3}m^2+\tfrac{2}{3}m\bar{\psi}_\mu\bar{\sigma}^{\mu\nu}\bar{\psi}_\nu+\tfrac{2}{3}m\psi_\mu\sigma^{\mu\nu}\psi_\nu\\
&+\Big\{\mathbf{\Lambda}_2-\mathbf{\Lambda}_2 Tr(\Gamma_{1}(\zeta))-\lambda_1^a\Big(\tfrac{\sqrt{2}}{2}\partial_\mu y^\mu_{a1}(\zeta)\\
&+\sqrt{2}\Big(\Gamma_{a2}(\zeta)
 +\tfrac{1}{3}m\zeta_a \black \Big) -\sqrt{2}m\zeta_a \black\Big)\\
&+\tfrac{\sqrt{2}}{2}i\lambda_1\sigma^\mu\bar{\psi}_\mu 
+\lambda_2\left(\partial_\mu t^\mu -\tfrac{1}{2}h\right)\\
&+\lambda_0\left(M^*  -m \left(\partial_\mu t^\mu  -\tfrac{1}{2}h\right) \right)  +h.c.\big\}\Big] \,,
\end{aligned}
\ee
where $\mathbf{D}_\mu$ is the covariant derivative on the background. The linearised local supersymmetry transformations are:
\be
\label{all_sugra_transforms_with_m} 
\begin{aligned}
\delta \psi_\mu&=-2\mathcal{D}_\mu\epsilon + \tfrac{i}{3}m\left(\varepsilon\sigma_\mu\bar{\epsilon}\right),\\
\delta^{(2)}\zeta&=-\epsilon\,,\\
\delta \lambda&=-\sqrt{2}\epsilon\mathbf{\Lambda_2}\,,\\
\delta  h_{\mu\nu}&=\delta M=\delta b_\mu=\delta t^\mu=\delta \lambda_0=\delta \lambda_2=0\,,
\end{aligned}
\ee 
where $\mathcal{D}_\mu\epsilon$ is defined in \eqref{cov_susy_transf} and the boundary conditions on the Lagrange multiplier superfield reduce to:
\be
\label{linear_boundary_cond}
\lambda_0\Big\vert_{\infty}=0,\quad \left(\lambda_1-\sqrt{2}\mathbf{\Lambda}_2\zeta\right)\Big\vert_{\infty}=0,\quad \lambda_2=0\Big\vert_{\infty}\,.
\ee
In order to study the gravitino mass we will need the equations of motion for the fermion fields in our theory. The gravitino e.o.m. is
\be
 \varepsilon^{\mu\nu\rho\sigma} \bar{\sigma}_\nu\mathbf{D}_\rho\psi_\sigma
+\tfrac{2}{3}m\bar{\sigma}^{\mu\nu}\bar{\psi}_\nu -i\tfrac{\sqrt{2}}{4} \bar{\sigma}^\mu  \lambda_1  
-i\tfrac{\mathbf{\Lambda_2}}{2}\bar{\sigma}^\mu\zeta=j^\mu\,,
\ee
where the source $\bar{j}_\mu$ accounts for the contributions from higher order terms in the perturbative expansions. Note that the local symmetry of the gravitino \eqref{all_sugra_transforms_with_m} implies the source conservation condition
\be
\label{source_conservation}
2\mathcal{D}^\mu\bar{j}_\mu+\tfrac{i}{3}m\bar{\sigma}^\mu j_\mu=0\,.
\ee
The equations for $\lambda_1$ and $\zeta$, respectively give:
\be
\label{zeta_lambda_1_eom}
\begin{aligned}
i\bar{\sigma}^\mu\psi_\mu&=2i\bar{\slashed{\mathbf{D}}}\zeta-\tfrac{4}{3}m\bar{\zeta}\,,\\
i\bar{\sigma}^\mu\psi_\mu&=\tfrac{\sqrt{2}}{\mathbf{\Lambda}_2}\left(i\bar{\slashed{\mathbf{D}}}\lambda_1-\tfrac{2}{3}m\bar{\lambda}_1 \right)\,.
\end{aligned} 
\ee
Taking the trace and the divergence of the gravitino equation, in conjunction with \eqref{zeta_lambda_1_eom} and \eqref{source_conservation} we obtain
\be
\begin{aligned}
i\bar{\sigma}^\mu\psi_\mu&=i\bar{\slashed{\mathbf{D}}}\left(\zeta+\tfrac{\sqrt{2}}{2\mathbf{\Lambda}_2}\lambda_1 \right)-\tfrac{2}{3}m\left(\zeta+\tfrac{\sqrt{2}}{2\mathbf{\Lambda}_2}\lambda_1 \right)\,,\\
-\tfrac{i}{4}m\sigma^\mu\bar{\psi}_\mu+\sigma^{\mu\nu}\mathbf{D}_\mu\psi_\nu &=\tfrac{\mathbf{\Lambda}_2}{2}\left(\zeta+\tfrac{\sqrt{2}}{2\mathbf{\Lambda}_2}\lambda_1\right)
+\tfrac{i}{4}\sigma^\mu\bar{j}_\mu\,.
\end{aligned}
\ee
Thus the effective goldstino at linear level is identified as the combination
\be 
\mathcal{G}\equiv \tfrac{1}{2}\left(\zeta+\tfrac{\sqrt{2}}{2\mathbf{\Lambda}_2}\lambda_1 \right)
\ee
We note that the orthogonal mode to $\mathcal{G}$
\be
\tau\equiv \tfrac{1}{2}\left(\zeta-\tfrac{\sqrt{2}}{2\mathbf{\Lambda}_2}\lambda_1 \right)
\ee
satisfies the equation
\be 
i\bar{\slashed{\mathbf{D}}}\tau-\tfrac{2}{3}m\bar{\tau}=0\,,
\ee
and is thus eliminated via the boundary conditions \eqref{linear_boundary_cond}. This is analogous the the elimination of the extra mode for the Weyl-Rarita Schwinger field in \cite{Blas:2008ce}. Finally, this allows us to recover  the standard equations for a massive gravitino coupled to a goldstino. The e.o.m. in our model are then equivalent to those arising from the perturbative action
\be 
\label{equivalent_action}
\begin{aligned}
\mathcal{L}=&\varepsilon^{\mu\nu\rho\sigma}\bar{\psi}_\mu\bar{\sigma}_\nu\mathbf{D}_\rho\psi_\sigma
+\tfrac{2}{3}m\bar{\psi}_\mu\bar{\sigma}^{\mu\nu}\bar{\psi}_\nu\\
&+2i\mathbf{\Lambda}_2\mathcal{G}\sigma^\mu\bar{\psi}_\mu-2i\mathbf{\Lambda}_2\mathcal{G}\slashed{\mathbf{D}}\bar{\mathcal{G}}+\tfrac{4}{3}m\mathbf{\Lambda}_2\mathcal{G}^2 +h.c.
\end{aligned}
\ee
In the flat limit, and taking $m\to0$ we recover the kinetic term of the Volkov-Akulov action for the goldstino, as expected (since we are working here in the linearised approximation).  We further note that supersymmetry is broken by the vev of $\mathbf{\Lambda}_2$. In the limit where it goes to zero,  supersymmetry is restored, as evidenced by the fact that the goldstino drops out of \eqref{equivalent_action}.  Of course, in this supersymmetric limit the backgrounds will be AdS, as revealed by equation \eqref{cc}.

\section{Conclusions} \label{sec:conc} 

In this paper, we developed a superspace version of unimodular gravity, where the chiral density superfield is constrained. This is similar to the way in which  the determinant of the metric is constrained in standard unimodular gravity, although it differs in that the constraint does not quite correspond to a local gauge fixing.  In any event, the constraint is most elegantly imposed using a superspace Lagrange multiplier in the form of  a chiral superfield.  By restoring general coordinate invaraince and local supersymmetry order by order using the St\"uckelberg trick, we have shown how the vev of the Lagrange multiplier can contribute a positive vacuum energy and allow for de Sitter vacua. More precisely, it is its top component that contributes to the vacuum energy, whilst at the same time spontaneously breaking supersymmetry.  Fluctuations about these vacua yield a massive gravitino coupled to a goldstino. The latter can be identified with a St\"uckelberg field introduced in the usual way to restore local supersymmetry. 

 At first glance, our work differs from previous formulations of pure de Sitter supergravity  which make use of a nilpotency constraint on a chiral superfield \cite{Bergshoeff:2015tra}.   There the nilpotency constraint is imposed directly using a Lagrange multiplier, motivated by the fact that the goldstino is expected to form part of a nilpotent superfield \cite{seiberg}.  The same ought to be true of the goldstino that emerges in our formulation, although it is not immediately obvious how this is actually realised.  We think a better understanding of the connection between the two formulations could enhance our understanding of both, perhaps leading to a unified description. 
 
 Our work can be extended in several other directions. The first of these is to develop a supersymmetric version of unimodular gravity in the Henneaux Teitelboim formulation \cite{HT}. This makes use of four-form field strengths rather than Jacobians of St\"uckelberg fields, although the two representations are closely related. A supersymmetric version of \cite{HT} could open up a natural path towards string theory realisation of our work, since four-forms are ubiquitous in flux compactifications \cite{Grana}.  A second motivation is to extend our ideas to a supersymmetric version of the sequestering proposal \citeseq\ for tackling the cosmological constant problem, again with a view towards a stringy embedding. 

\acknowledgements  We would like to thank Graham Shore, David Stefanyszyn and Gianmassimo Tasinato for useful discussions. The work of S.~N. and A.~P.~was supported by a Leverhulme Research Project Grant. The work of A.~P.~is also supported by an STFC Consolidated Grant.   I.~Z.~is partially supported by STFC, grant ST/P00055X/1.

\appendix
\section{Conventions and useful quantities}\label{Conventions and useful quantities}
We are using the conventions of \cite{Wess:1992cp}. We work with the mostly plus signature metric $\eta_{\mu\nu}=\text{diag}(-1,1,1,1)$ and 
\be
\varepsilon_{21}=\varepsilon^{12}=1,\quad \varepsilon_{12}=\varepsilon^{21}=-1,\quad \varepsilon_{11}=\varepsilon_{22}=0 \,.
\ee
The Pauli matrices are
\be
\label{pauli_matrices}
\sigma^0=
\bigl(\begin{smallmatrix}
-1 &0 \\ 
0 &-1 
\end{smallmatrix}\bigr) ,\quad
\sigma^1=
\bigl(\begin{smallmatrix}
0 & 1\\ 
1 & 0
\end{smallmatrix}\bigr),\quad
\sigma^2=
\bigl(\begin{smallmatrix}
0 & -i\\ 
i & 0
\end{smallmatrix}\bigr), \quad
\sigma^3=
\bigl(\begin{smallmatrix}
1 & 0\\ 
0 & -1
\end{smallmatrix}\bigr)
\ee
and 
\be
\bar{\sigma}^0=\sigma^0,\quad \bar{\sigma}^1=-\sigma^{1,2,3} \,.
\ee
The chiral Supergravity superfield in component form is:
\be
\begin{aligned}
R=&-\tfrac{1}{6}\Big\{M+\Theta\left[\sigma^\mu\bar{\sigma}^\nu\psi_{\mu\nu}-i\sigma^\mu\bar{\psi}_\mu M+i\psi_\mu b^\mu \right]\\
&+\Theta^2\Big[-\tfrac{1}{2}\mathcal{R}+i\bar{\psi}^\mu\bar{\sigma}^\nu\psi_{\mu\nu}+\tfrac{2}{3}MM^*
+\tfrac{1}{3}b^\mu b_\mu\\
&-ie_a^\mu\mathcal{D}_\mu b^a+\tfrac{1}{2}\bar{\psi}\bar{\psi}M-\tfrac{1}{2}\psi_\mu\sigma^\mu\bar{\psi}_\nu b^\nu\\
&+\tfrac{1}{8}\varepsilon^{\mu\nu\rho\sigma}\left[\bar{\psi}_\mu\bar{\sigma}_\nu\psi_{\rho\sigma}+\psi_\mu\sigma_\nu\bar{\psi}_{\rho\sigma}\right]
\Big]\Big\}\,,
\end{aligned} 
\ee
with
\be
\psi_{\mu\nu}=2\tilde{\mathcal{D}}_{[\mu}\psi_{\nu]},\quad \tilde{\mathcal{D}}_\mu\psi_\nu=\partial_\mu\psi_\nu+\psi_\nu\omega_\mu  \,.
\ee

\section{Supergravity Transformations}\label{app:Supergravity Transformations}
The susy transformations of the supergravity fields (up to linear order in $\epsilon$) are
\be
\label{all_sugra_transforms} 
\begin{aligned}
\delta  e_\mu^{\mathbf{a}}=&i\left(\psi_\mu\sigma^{\mathbf{a}}\bar{\epsilon}-\epsilon\sigma^{\mathbf{a}}\bar{\psi}_\mu\right)\,,\\
\delta \psi_\mu^m=&-2\mathcal{D}_\mu\epsilon^m +\tfrac{i}{3}M\left(\varepsilon\sigma_\mu\bar{\epsilon}\right)^m+ib_\mu\epsilon^m+\tfrac{i}{3}b^\rho\left(\epsilon\sigma_\rho\bar{\sigma}_\mu\right)^m, \\
\delta M=&-\epsilon\left(\sigma^\alpha\bar{\sigma}^\beta\psi_{\alpha\beta}+ib^\alpha\psi_\alpha-i\sigma^\alpha\bar{\psi}_\alpha M \right)\,,\\
\delta b_{\alpha\dot{\alpha}}=&\epsilon\mathcal{F}(\psi;M,b)\\
=&\epsilon^\delta\Big[\tfrac{3}{4}\bar{\psi}_{\alpha\phantom{\gamma}\delta\dot{\gamma}\dot{\alpha}}^{\phantom{\alpha}\dot{\gamma}}+\tfrac{1}{4}\epsilon_{\delta\alpha}\bar{\psi}^{\gamma\dot{\gamma}}_{\phantom{\alpha\beta}\gamma\dot{\alpha}\dot{\gamma}}-\tfrac{i}{2}M^*\psi_{\alpha\dot{\alpha}\delta}\\
&\quad\quad+\tfrac{i}{4}\Big(\bar{\psi}_{\alpha\dot{\rho}}^{\phantom{\alpha\beta}\dot{\rho}}b_{\delta\dot{\alpha}}
+\bar{\psi}_{\delta\dot{\rho}}^{\phantom{\alpha\beta}\dot{\rho}}b_{\alpha\dot{\alpha}}-\bar{\psi}_{\delta\phantom{\alpha}\dot{\alpha}}^{\phantom{\alpha}\dot{\rho}}b_{\alpha\dot{\rho}}\Big)\Big]\\
&-\bar{\epsilon}\Big[\tfrac{3}{4}\psi^\gamma_{\phantom{\alpha}\dot{\delta}\gamma\dot{\alpha}\alpha}
+\tfrac{1}{4}\varepsilon_{\dot{\delta}\dot{\alpha}}\psi_{\alpha\phantom{\alpha\beta}\dot{\gamma}\gamma}^{\phantom{\alpha}\dot{\gamma}\gamma}+\tfrac{i}{2}M\bar{\psi}_{\alpha\dot{\alpha}\dot{\delta}}\\
&-\tfrac{i}{4}\Big(\psi_{\rho\dot{\alpha}}^{\phantom{\alpha\beta}\rho}b_{\alpha\dot{\delta}}
+\psi_{\rho\dot{\delta}}^{\phantom{\alpha\beta}\rho}b_{\alpha\dot{\alpha}}-\psi_{\phantom{\alpha}\dot{\delta}\alpha}^\rho b_{\rho\dot{\alpha}}\big)\Big]\,,
\end{aligned}
\ee
where space-time indices have been converted to spinor indices through contraction with the Pauli matrices \eqref{pauli_matrices} and:
\be
\label{cov_susy_transf}
\mathcal{D} _\mu\epsilon^m=\partial_\mu\epsilon^m+\epsilon^b\omega_{\mu b}^{\ \ m} \,.
\ee
Working to second order in the fermions, the supergravity transformations of the quantities defined in \eqref{y_gamma_notation} are then:
\be
\begin{aligned}
\delta^{(s)}[y^{\mu}_{1}(\epsilon)]=& -  \xi^\rho \partial_\rho \Big[ 2i\sigma^\mu\Big]\bar{\epsilon} + \partial_\rho \xi^\mu y^{\rho}_{1}(\epsilon)\,,\\
\delta^{(s)}[y^\mu_2(\epsilon)]=&-\tfrac{1}{2}y^{\nu}_1(\epsilon) D_\nu y^\mu_{1}(\epsilon)+\Gamma_2(\epsilon)\black y^\mu_{1}(\epsilon)\\
&-\xi^\rho\partial_\rho \Big[ \bar{\psi}_\nu\bar{\sigma}^\mu\sigma^\nu\Big]\bar{\epsilon}           
+\partial_\rho \xi^\mu y^\rho_2(\epsilon) - \tfrac{4i}{3}M^*\epsilon\sigma^\mu\bar{\epsilon}\\
& -ib_\nu\bar{\epsilon}\bar{\sigma}^\mu\sigma^\nu\bar{\epsilon}+i b_\nu\bar{\epsilon}\bar{\sigma}^\nu\sigma^\mu\bar{\epsilon}\,, \\
\delta^{(s)}[Tr(\Gamma_{1}(\epsilon))]=&-\partial_\mu\epsilon y_{1}^\mu(\epsilon)+2\epsilon\Gamma_2(\epsilon)
  -  \xi^\rho\partial_\rho\Big[ i\psi_\mu\sigma^\mu\Big]\bar{\epsilon} \\
&-\tfrac{2}{3}M^*\epsilon^2 -\tfrac{4}{3}M\bar{\epsilon}^2 \,,\\
\delta^{(s)}[\Gamma_2(\epsilon)]=& -  \xi^\rho\partial_\rho \Big[ -i\omega_\mu\sigma^\mu  \Big]\bar{\epsilon}  - \xi^\rho\partial_\rho M^* \tfrac{1}{3}\epsilon \\
& -\tfrac{1}{6}\xi^\rho\partial_\rho\Big[ b_\mu\varepsilon\sigma^\mu\Big]\bar{\epsilon}\,.
\end{aligned}
\ee

A chiral density superfield $\Delta$ is defined by its transformation law
\be
\delta\Delta=-\partial_N[(-1)^N\eta^N] \,,
\ee 
with $\eta^\mu,\eta^\alpha$ defined in \eqref{etas_def}. In components
\be
\Delta=a+\sqrt{2}\Theta\rho+\Theta^2 f \,,
\ee
and then the transformation rules of the components will be
\be 
\label{density_comps_tranform}
\begin{aligned}
\delta a=&-\sqrt{2}\epsilon\rho+ia\psi_\mu\sigma^\mu\bar{\epsilon}\,,\\
\delta\rho=&-\sqrt{2}\epsilon f -i\sqrt{2}\mathcal{D}_\mu\left(\sigma^\mu\bar{\epsilon}a\right)+i\psi_\mu\sigma^\mu\bar{\epsilon}\rho\\
&+i\left(\sigma^\mu\bar{\epsilon}\right) \psi_\mu\rho-\tfrac{\sqrt{2}}{3}\epsilon M^* a\\
&-\tfrac{\sqrt{2}}{6}a\sigma^\mu\bar{\epsilon}b_\mu+\tfrac{\sqrt{2}}{2}\psi_\nu\bar{\psi}_\mu\bar{\sigma}^\nu\sigma^\mu\bar{\epsilon}a\,,\\
\delta f=&\partial_\mu\left[-a\bar{\psi}_\nu\bar{\sigma}^\mu\sigma^\nu\bar{\epsilon}+i\sqrt{2}\rho\sigma^\mu\bar{\epsilon} \right]\,.
\end{aligned}
\ee

\bibliography{refs}
\bibliographystyle{utphys}

\end{document}